\documentclass[
    aps,
    pra,
    nofootinbib,
    longbibliography,
    twocolumn,
    superscriptaddress
]{revtex4-2}

\usepackage{graphicx}
\usepackage{dcolumn}
\usepackage{bm}
\usepackage{subfigure}
\usepackage{qcircuit}
\usepackage{amssymb}
\usepackage{amsmath}
\usepackage{mathtools}
\usepackage{dsfont}
\usepackage{xcolor}
\usepackage{qcircuit}
\usepackage{hyperref}
\usepackage{physics}
\hypersetup{
    colorlinks=true,
    citecolor=blue
}
\usepackage{placeins}
\usepackage{float}

\begin{document}

\title{Training Quantum Boltzmann Machines with the $\beta$-Variational Quantum Eigensolver}

\author{Onno Huijgen}
\email{onno.huijgen@donders.ru.nl}
\affiliation{Radboud University, Nijmegen, The Netherlands}

\author{Luuk Coopmans}
\email{luuk.coopmans@quantinuum.com}
\affiliation{Quantinuum, Partnership House, Carlisle Place, London SW1P 1BX, United Kingdom}

\author{Peyman Najafi}
\affiliation{Radboud University, Nijmegen, The Netherlands}

\author{Marcello Benedetti}
\affiliation{Quantinuum, Partnership House, Carlisle Place, London SW1P 1BX, United Kingdom}

\author{Hilbert J. Kappen}
\affiliation{Radboud University, Nijmegen, The Netherlands}

\date{February 27, 2024}

\begin{abstract}
\nocite{Lloyd_2016,Zhao_2019, Cerezo_2021,Benedetti_2019,Tilly_2022,Amin_2018,Benedetti_2017,Kieferova_2017,Kappen_2020,Bravyi_2021,Hinton_2002}
The quantum Boltzmann machine (QBM) is a generative machine learning model for both classical data and quantum states. Training the QBM consists of minimizing the relative entropy from the model to the target state. This requires QBM expectation values which are computationally intractable for large models in general. It is therefore important to develop heuristic training methods that work well in practice. In this work, we study a heuristic method characterized by a nested loop: the inner loop trains the $\beta$-variational quantum eigensolver ($\beta$-VQE) by Liu et al.~\cite{Liu_2021} to approximate the QBM expectation values; the outer loop trains the QBM to minimize the relative entropy to the target. We show that low-rank representations obtained by $\beta$-VQE provide an efficient way to learn low-rank target states, such as classical data and low-temperature quantum tomography. We test the method on both classical and quantum target data with numerical simulations of up to 10 qubits. For the cases considered here, the obtained QBMs can model the target to high fidelity. We implement a trained model on a physical quantum device. The approach offers a valuable route towards variationally training QBMs on near-term quantum devices.
\end{abstract}

\maketitle

\section{Introduction}
The study of quantum information technologies has shown to be a fruitful discipline for the discovery of novel algorithms and computing applications. It is expected that quantum algorithms will outperform their classical counterparts on several tasks of practical relevance, such as factoring large prime numbers and simulating quantum many-body systems.  Over the last decade, scientists have been exploring if quantum algorithms can also offer advantages in machine learning. While this has been answered in the affirmative~\cite{Lloyd_2016,Zhao_2019}, the theoretical performance benefits will materialize only when large-scale fault-tolerant quantum computers become available.

In the context of near-term quantum computing, a plethora of variational quantum algorithms (VQAs) for machine learning tasks have been investigated~\cite{Cerezo_2021,Benedetti_2019,Tilly_2022}.  In VQAs, one solves the computational problem by choosing an ansatz, typically a parameterized quantum circuit, and then optimizes its parameters with respect to a suitable objective function. Applications of VQAs include finding ground states of small molecules, combinatorial optimization, classification, and generative modeling.

In this paper, we continue this line of research and propose a VQA for training quantum Boltzmann machines (QBMs)~\cite{Amin_2018,Benedetti_2017}. QBMs are generative models inspired by quantum Ising models and can be used for both classical and quantum data. In QBM training, the goal is to find the weights of a Hamiltonian ansatz, such that its thermal state best approximates the target density matrix~\cite{Kieferova_2017,Kappen_2020}. The QBM training depends on the ability to prepare and compute properties of a thermal state, which is intractable in general~\cite{Bravyi_2021}. It is therefore key to develop efficient heuristic methods that work well in practice; indeed, even classical Boltzmann machines are intractable and must be trained by efficient heuristics such as contrastive divergence~\cite{Hinton_2002}. In this work, we employ the recently proposed $\beta$-variational quantum eigensolver~\cite{Liu_2021} ($\beta$-VQE) to approximate the QBM properties.

$\beta$-VQE represents a mixed density matrix as a combination of a classical probability distribution and a parameterized quantum circuit. Computational basis states are randomly sampled and used as input for the circuit, where they are transformed into quantum states. $\beta$-VQE has been experimentally demonstrated on a superconducting quantum computer~\cite{Guo_2023}. The purpose of our work is to provide evidence that $\beta$-VQE can be successfully exploited for the training of QBMs on classical and quantum data sets to high accuracy. The algorithm is illustrated in Fig.~\ref{fig:dl} along with some empirical statevector simulation results on a classical data set. 

Several other techniques have been proposed for QBM training: quantum annealing~\cite{Benedetti_2017}, variational imaginary time evolution~\cite{Zoufal_2021}, the eigenstate thermalization hypothesis~\cite{Anschuetz_2019}, pure thermal shadows~\cite{Coopmans_2023}, and others~\cite{Wiebe_2019,Kieferova_2021}. Most of these approaches have not been demonstrated as they require a fault-tolerant quantum computer~\cite{Anschuetz_2019,Coopmans_2023,Wiebe_2019,Kieferova_2021}. On the other hand, the few approaches that have been demonstrated on real hardware~\cite{Benedetti_2017,Zoufal_2021} incur a significant overhead in terms of ancilla qubits. Remarkably, the approach studied here is near-term and uses the same number of qubits as the number of variables in the data set. We deploy our trained model by preparing and sampling it on IBM Kolkata device.

We equip the algorithm with additional features and rules of thumb that reduce the overall computational cost. Firstly, it is possible to use a limited number of computational basis states in $\beta$-VQE, yielding a low-rank approximation. We provide evidence that this is especially useful for classical target data. Secondly, since small updates in parameter space during QBM training lead to small steps in density matrix space, we introduce a natural warm-start strategy for $\beta$-VQE. We observe that warm-start yields faster convergence, possibly avoiding barren plateaus~\cite{McClean_2018} by keeping the solution close to a local minimum at all times. Finally, we find that a circuit depth scaling linearly with the system size is sufficient to get accurate results. For the dataset we use, the $\beta$-VQE approximation resulted in a difference in fidelity of $\mathcal{O}(10^{-4})$ compared QBM training using the exact gradient, regardless of system size, suggesting that this method is scalable.

\section{Methods}
\label{sec:methods}

\subsection{Quantum Boltzmann machines}
\label{sec:QBM_rev}

The QBM~\cite{Amin_2018,Benedetti_2017} is defined as the $n$-qubit Gibbs state (mixed density matrix)
\begin{equation}
\label{eq:sigma}
\sigma_w = \frac{e^{-\beta H_w}}{Z},
\end{equation}
with $Z = \Tr(e^{-\beta H_w})$ the partition function, $\beta > 0$ the fixed inverse temperature and $H_w$ a parameterized Hamiltonian. In this work, we consider Hamiltonians of the form 
\begin{equation}
H_w = \sum_{r} w_r H_r,
\end{equation} which consists of a sum of a polynomial number of non-commuting qubit-operators $H_r$ with weights $w_r$. The specific form of the operators $H_r$ does not matter at this point and will be given in the result section below. 
The quantum Hamiltonian $H_w$ generalizes the role of the energy function in a classical Boltzmann machine~\cite{Ackley_1985}.

In order to find the optimal weights $w^*$, we aim to minimize the quantum relative entropy
\begin{equation}
  \label{eq:QuantumRelent}
  S(\eta \| \sigma_w) = \Tr(\eta \log{\eta}) - \Tr(\eta \log{\sigma_w})
\end{equation} 
from $\sigma_w$ to the target density matrix $\eta$~\cite{Kieferova_2017,Kappen_2020}. This information-theoretic measure generalizes the classical Kullback-Leibler divergence to density matrices and has nice properties for the QBM model ansatz, e.g., the logarithm cancels the exponential in Eq.~\eqref{eq:sigma}. 
The target $\eta$ is either some quantum system with unknown Hamiltonian $H_t$ at finite temperature, i.e., $\eta = \frac{e^{-\beta_t H_t}}{Z_t}$, or a classical data set \emph{embedded} into a quantum state. 

The optimization of $S(\eta \|\sigma_w)$ can be done with gradient descent. By noticing that only the second term in Eq.~\eqref{eq:QuantumRelent} depends on $w_r$, and using Duhamel's formula for the derivative of a matrix exponential~\cite{Haber_2018}, we obtain the partial derivatives
\begin{equation}
\label{eq:grad}
\frac{\partial S}{\partial w_r} = \tr(H_r \eta) - \tr(H_r \sigma_w) .
\end{equation}
Similar to the classical case~\cite{Ackley_1985}, the gradient is given by the difference between statistics under the data density matrix $\eta$ and the model density matrix $\sigma_w$. The data statistics do not change during training and need to be obtained only once. However, the QBM statistics change after every iteration of gradient descent and must be recomputed. The QBM density matrix $\sigma_w$ in Eq.~\eqref{eq:sigma} is given by the matrix exponential of the Hamiltonian, whose dimension scales exponentially in the number of qubits. This means that computing the full density matrix becomes computationally intractable for larger system sizes.

While the classical BM has the same exponential scaling of the number of states, there exist several approximation algorithms that can effectively compute the necessary statistics, often based on Markov-chain Monte Carlo~\cite{Zhang_2018}. However, for quantum systems, no general MCMC methods exist due to the negative sign problem for non-stoquastic models~\cite{Loh_1990,Pan_2022}. Since the QBM is not restricted to the stoquastic domain and can include spin-glass models, quantum Monte Carlo is not an option in general. In the following section, we review the $\beta$-VQE method that with help of a quantum computer can be used to potentially circumvent this issue.

\subsection{Classical data embedding}
\label{sec:data_prep}
The QBM can be trained on classical and quantum data. Since classical data is not naturally represented in density matrices, the data is embedded in a quantum system. There are various ways to do this. We use the pure-state embedding presented in~\cite{Kappen_2020}. From a classical data set $\mathcal{D}$, the empirical probability distribution is constructed counting the relative occurrence of each binary vector in the data, $q(\bm{s}^{i})=\frac{1}{|\mathcal{D}|} \sum_{\bm{s}^j \in \mathcal{D}} \delta_{\bm{s}^{i},\bm{s}^j}$. We create a pure quantum state representation from the empirical distribution by mapping the $n$-dimensional binary vector $\bm{s}^i$ to a spin-$\frac{1}{2}$ system with $n$ particles. Each $1$ in $\bm{s}^i$ corresponds with a particle in the spin-up state $\ket{1}$ and each $0$ to a particle in spin-down state $\ket{0}$ in the $\sigma^z$ basis. Let:
\begin{equation}
    \label{eq:notation}
    \ket{\bm{s}^i} = \bigotimes_{j=0}^{n} \ket{s^i_j}.
\end{equation}
The quantum state is obtained by taking the superposition of all spin states corresponding to the binary vectors with coefficients equal to $\sqrt{q(\bm{s})}$, i.e. $\ket{\psi} = \sum_{\bm{s} \in \mathcal{D}} \sqrt{q(\bf{s})} \ket{\bm{s}}$. The density matrix embedding is then constructed by taking the outer product of this state,
\begin{equation}
    \label{empirical}
    \eta = \outerproduct{\psi}{\psi} = \sum_{\bm{s}, \bm{s}^\prime} \sqrt{q(\mathbf{\bm{s}})q(\mathbf{\bm{s}^\prime})}\dyad{\bm{s}}{\bm{s}^\prime}.
\end{equation}

\subsection{\texorpdfstring{$\beta$}{beta}-VQE}

Recently, Liu et al.~\cite{Liu_2021} proposed a variational method to represent a mixed density matrix called $\beta$-VQE. It combines a classical network with a quantum circuit~(Fig.~\ref{fig:convergence_salamander}) into a parameterized density matrix ansatz given by
\begin{equation}
\label{eq:rho}
\rho_{\theta, \phi} = \sum_{\mathbf{s}}p_{\phi}(\mathbf{s}) U_{\mathbf{\theta}} \outerproduct{\mathbf{s}}{\mathbf{s}}U^{\dagger}_{\mathbf{\theta}}.
\end{equation} 
Here $p_{\phi}(\mathbf{s})$ is a parameterized probability distribution over binary states implemented by a classical network, e.g., a parameterized Bernoulli distribution or an autoregressive network~\cite{Germain_2015}. The binary states are then used as inputs for a parameterized quantum circuit $U_\theta$, which transforms them into quantum states.
The combined system $\rho_{\theta, \phi}$ can accurately represent finite-temperature Gibbs states of quantum spin systems, e.g., see Ref.~\cite{Liu_2021}. Here we aim to see if it can be used for the wide range of Hamiltonians needed to train the QBM on quantum and classical data. 

The optimal parameters of $\rho_{\theta, \phi}$ for a fixed QBM Hamiltonian $H_w$ are found by minimizing the variational free energy 
\begin{equation}
  \label{eq:free_energy}
  \mathcal{F}(\sigma_w, \rho_{\theta, \phi}) = \Tr(\rho_{\theta, \phi} \log{\rho_{\theta, \phi}}) + \Tr(\rho_{\theta, \phi} H_w). 
\end{equation}
This is equivalent to minimizing the quantum relative entropy from $\sigma_w$ to $\rho_{\theta, \phi}$.
The minimization can be done with gradient descent on the classical network and quantum circuit parameters simultaneously. The required gradients~\cite{Liu_2021} are given by
\begin{align}    
\label{eq:GradTheta}
    \nabla_{\theta} \mathcal{F} & = \sum_{\mathbf{s}} p_{\phi}(\mathbf{s}) \left[ \nabla_{\theta} \expval{U^{\dagger}_{\theta} H_{w} U_{\theta}}{\mathbf{s}} \right],
\end{align} 
for the circuit, and 
\begin{align}
\label{eq:GradPhi}
    \nabla_{\phi} \mathcal{F} & = \sum_{\mathbf{s}} p_{\phi}(\mathbf{s}) \left[ (f(\mathbf{s}) - b) \nabla_{\phi} \log{p_{\phi}(\mathbf{s})} \right],
\end{align}
for the classical network. Here we used a control variate to reduce the variance as in~\cite{Liu_2021}: $f(\mathbf{s}) = \ln{p_{\phi}(\mathbf{s})} + \expval{U_\theta^\dag H_w U_\theta}{\mathbf{s}}$ and $b = \sum_{\mathbf{s}} p_{\phi}(\mathbf{s}) f(\mathbf{s})_{p_{\phi}}$.
For quantum gates of the form $R_G(\theta_j) = e^{- i \theta_j G /2}$ where $G$ is a Pauli operator, the gradient w.r.t. the circuit parameters, Eq.~\eqref{eq:GradTheta}, reduces to a parameter shift rule: a weighted sum of energy expectation values~\cite{Mitarai_2018}. Suppose the circuit is a product of $L$ such gates, i.e. $U_{\theta} = R_{G_L}(\theta_L) R_{G_{(L-1)}}(\theta_{L-1}) \cdots R_{G_1}(\theta_1)$. Then
\begin{multline}
    \label{eq:ParameterShift}
    \frac{\partial}{\partial \theta_i} \expval{U^{\dagger}_{\theta} H_{w} U_{\theta}}{\mathbf{s}} = \\
    \frac{1}{2} \Big[ \expval{R^{\dagger}_{G_i}(\theta_i + \tfrac{\pi}{2}) \tilde{H}_w R_{G_i}(\theta_i + \tfrac{\pi}{2})}{\psi_{i-1}} - \\ 
    \expval{R^{\dagger}_{G_i}(\theta_i - \tfrac{\pi}{2}) \tilde{H}_w R_{G_i}(\theta_i - \tfrac{\pi}{2})}{\psi_{i-1}} \Big] ,
\end{multline}
where for convenience, $\ket{\psi_{i-1}} = \prod_{j=1}^{i-1} R_{G_j}(\theta_j) \ket{\mathbf{s}}$ and $\tilde{H}_w = \prod_{j=(i+1)}^{L} R^{\dagger}_{G_j}(\theta_j) H_w \prod_{j=L}^{i+1} R_{G_j}(\theta_j)$.

Computation of the gradient is tractable on a quantum computer if the sum is approximated by sampling a polynomial number of bitstrings from  $p_\phi(\mathbf{s})$, and if there is no barren plateau in the optimization landscape~\cite{McClean_2018}. The gradient w.r.t. the parameters of the classical network, Eq.~\eqref{eq:GradPhi}, has the form of a standard weighted entropy gradient, which for classical networks can be efficiently approximated from sampling~\cite{Shakir_2019}. 

\subsection{Truncated-rank \texorpdfstring{$\beta$}{beta}-VQE}

We propose a slight modification of the sampling approach for the gradients originally presented in~\cite{Liu_2021}. Instead of randomly sampling from all computational basis states, we choose the $R$ states with the highest probability and renormalize $p_{\phi}$ on those states. This results in a $\beta$-VQE density matrix of rank $R$ 
\begin{equation}
  \label{eq:rho_rank}
  \rho_{\theta, \phi} = \frac{1}{\sum_{k=i_1}^{i_R} p_{\phi}(\bm{s}_k)}  \sum_{j=i_1}^{i_{R}} p_{\phi}(\bm{s}_{j}) U_\theta \dyad{\bm{s}_j}{\bm{s}_j} U_\theta^\dag.
\end{equation}
In the following, we refer to this as truncated-rank $\beta$-VQE. As an immediate consequence, the gradients for the variational free energy Eqs.~\eqref{eq:GradTheta} and~\eqref{eq:GradPhi} are also truncated for this model. This means that we can heuristically choose a small $R$ so that the optimization can be performed at a reduced computational cost.

\subsection{Training QBMs with the nested-loop algorithm}

\begin{figure}[ht!]
    \centering
    \hspace{1cm}
    \includegraphics[width=8cm]{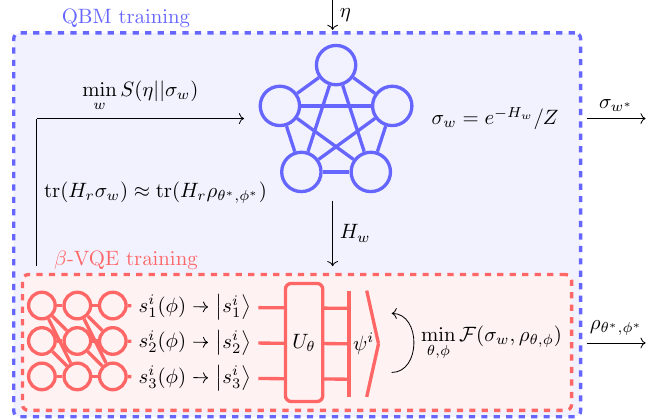} 
    \caption{Illustration of the nested-loop algorithm for QBM training. In the outer loop (blue box) we optimize the quantum relative entropy $S$ from the QBM model $\sigma_w$ to the target $\eta$. This step utilizes an approximation to the QBM statistics, $\tr(H_r \sigma_w) \approx \tr(H_r \rho_{\theta^*,\phi^*})$, which is provided by the inner loop (red box). In the inner loop we optimize the variational free energy $\mathcal{F}$ of the $\beta$-VQE ansatz $\rho_{\theta, \phi}$ with the QBM Hamiltonian $H_{w}$. At the end of training, we output the QBM model and its $\beta$-VQE approximation.
    }
    \label{fig:dl}
\end{figure}

We now describe the nested-loop algorithm to train a QBM with a truncated rank $\beta$-VQE. Recall that the aim is to train the QBM so that it resembles some target data embedded in $\eta$.
This is schematically shown in Figure~\ref{fig:dl}. The algorithm starts with a simple ansatz for the QBM Hamiltonian $H_w$, e.g., a Heisenberg XXZ model or a random spin-glass model. In the \emph{inner loop}, the $\beta$-VQE ansatz $\rho_{\theta, \phi}$ is trained to represent the QBM $\sigma_w$ by minimizing Eq.~\eqref{eq:free_energy}. In the \emph{outer loop}, $\rho_{\theta, \phi}$ is used to compute approximate QBM statistics $\tr( H_r \sigma_w) \approx \tr(H_r \rho_{\theta^*,\phi^*})$. These are used for training the QBM by gradient descent with Eq.~\eqref{eq:grad}.

Since $H_w$ changes only slightly between two QBM iterations, we use a warm-start strategy for the parameters $\theta, \phi$ of the $\beta$-VQE ansatz $\rho_{\theta, \phi}$ between the steps of the outer loop. We reuse the converged parameters of $\rho_{\theta, \phi}$ in one $\beta$-VQE inner loop as the initial parameters for the next $\beta$-VQE inner loop. Hence, while the $\beta$-VQE has to run for a relatively long time in the first inner loop, we find that it converges much faster in subsequent iterations. 

The total number of statistic estimations of the nested-loop algorithm is of order
\begin{equation}
\label{eq:eval}
  \mathcal{O}( n_{\text{QBM}} \times n_{\beta\text{-VQE}} \times R \times n_{\theta} ).
\end{equation}
Here $n_{\text{QBM}}$ is the number of QBM iterations for convergence to some fixed accuracy, $n_{\beta\text{-VQE}}$ the number of $\beta$-VQE iterations, $R$ the rank of the $\beta$-VQE, and $n_{\theta}$ the number of circuit parameters.

\begin{figure*}[t]
\hspace{-12cm} (a) \hspace{5cm} (b)\\
    \subfigure{
\Qcircuit @C=.5em @R=1.0em {
 & \multigate{1}{SU(4)} & \qw & \sgate{SU(4)}{3} & \qw\\
 & \ghost{SU(4)} & \multigate{1}{SU(4)} & \qw & \qw\\
 & \multigate{1}{SU(4)} & \ghost{SU(4)} & \qw & \qw\\
 & \ghost{SU(4)} & \qw & \gate{SU(4)} & \qw
}
\hspace{5.7cm}
\raisebox{-0.0em}{
\subfigure{
\Qcircuit @C=1.0em @R=0.7em {
  & \multigate{1}{SU(4)} & \qw \\
  & \ghost{SU(4)} & \qw
} }}
\hspace{0.3cm}
\raisebox{-1.0em}{=}
\hspace{0.3cm}
\raisebox{-4.5em}{
\hspace{-8.4cm}
\Qcircuit @C=.5em @R=.7em {
  & \gate{R_z(\theta_1)} & \gate{R_y(\theta_3)} & \gate{R_z(\theta_5)} & \ctrl{1} & \gate{R_z(\theta_7)} & \targ & \qw & \ctrl{1} & \gate{R_z(\theta_{10})} & \gate{R_y(\theta_{12})} & \gate{R_z(\theta_{14})} &  \qw \\
  & \gate{R_z(\theta_2)} & \gate{R_y(\theta_4)} & \gate{R_z(\theta_6)} & \targ & \gate{R_y(\theta_8)} & \ctrl{-1} & \gate{R_y(\theta_9)} & \targ & \gate{R_z(\theta_{11})} & \gate{R_y(\theta_{13})} & \gate{R_z(\theta_{15})} & \qw 
}}
}
\caption{Parameterized quantum circuit used for the $\beta$-VQE. (a) shows a single layer of a four-qubit ansatz. It consists of a checkerboard pattern of general $SU(4)$ unitaries with periodic boundary conditions. (b) The $SU(4)$ unitaries are implemented as three CNOT gates and 15 single-qubit rotations. This is a special case of the decomposition given in Figure~2 of ~\cite{Shende_2004}.}
\label{fig:circuit}
\end{figure*}

\section{Numerical Results}~\label{sec:results}

Using the nested-loop algorithm, we train QBMs on both classical and quantum data. We use the Hamiltonian ansatz
\begin{equation}
    \label{eq:H_qbm}
    H_w = \sum_{k \in \{x, z\}} \sum_{i=1}^n w^{k}_{i} \sigma_i^k + \sum_{k \in \{x, y, z \}}\sum_{i<j} \tilde{w}_{i,j}^k \sigma_i^k \sigma_j^k,
\end{equation}
where $w$ and $\tilde{w}$ are weights, and $\sigma_i^k$ is the $n$-qubit Pauli operator acting on qubit $i$. We initialize the weights randomly from a normal distribution with mean $0$ and standard deviation $\frac{1}{\sqrt{n}}$. The weights are trained in the outer loop using gradient descent with a momentum parameter $\alpha = \frac{1}{2}$~\footnote{While the specifics of the update scheme do not noticeably change the quality of the final result, we find that they do have a large impact on the number of outer loop iterations needed to get there. Fast convergence was usually achieved with a variable learning rate that is increased by $1\%{}$ after every iteration as long as the cost function decreased, and halved if the cost function increased, combined with some momentum.}.
The ansatz taken for $H_w$ determines the precise models that the QBM can represent. Not every state in the Hilbert space can be represented by this ansatz. Therefore, there might be target data that can not be exactly modeled. We define this as a model mismatch.

For the $\beta$-VQE, we implement the classical network $p_{\phi}$ with a variational auto-encoder (VAE)~\cite{Germain_2015} made of a hidden layer of $50$ neurons. For the quantum circuit $U_\theta$, we use a checkerboard pattern of general $SU(4)$ unitaries with periodic boundary conditions. One layer of this ansatz is shown in Fig.~\ref{fig:circuit}(a). Each $SU(4)$ unitary is implemented via a decomposition into CNOT gates and single-qubit rotations following~\cite{Shende_2004}. This is illustrated in Fig.~\ref{fig:circuit}(b).
We define the circuit depth $d$ as the number of repetitions of such a layer. This gives a total number of circuit parameters $n_\theta=15 \times d \times n$. We determine the optimizer settings, e.g., the learning rate, for each target data set separately, and give them below. As the stopping criterion for the inner loop, we use a target precision $\epsilon$ on the $\beta$-VQE gradient (Eqs.~\eqref{eq:GradTheta} and~\eqref{eq:GradPhi}). In addition, we also set a maximum number of iterations $n_{\beta_{\mathrm{VQE}}}^{\mathrm{max}}=2000$ for the inner loop, in order to avoid very long convergence times. 
To quantify the model performance, we compute the Uhlmann-Jozsa fidelity between the QBM and the target state, given by
\begin{equation}
    \label{eq:fidelity}
    F(\eta, \rho) = \Tr(\sqrt{\sqrt{\eta}\rho\sqrt{\eta}})^2.
\end{equation}
This fidelity satisfies all of Jozsa's axioms, described in~\cite{Liang_2019}. Importantly, fidelity is an intensive property, as opposed to the quantum relative entropy. Thus it is a useful measure for comparing performance on different system sizes.

By means of classical statevector simulations, we show that with these settings we are able to train QBMs to high accuracy on both classical and quantum data. Afterward, we train a small QBM using noisy and noise-free quantum simulators of real quantum hardware and sample the final QBM on the real device.

\subsection{Classical data results}
The classical data we consider in this paper is introduced in~\cite{Berry_2017}. A patch of 160 connected ganglion cells from the salamander retina is connected to a multi-electrode array, where the activity of multiple neurons is recorded simultaneously. The ganglion cells are then exposed to 297 repeats of a movie. Every timeframe of $20$ ms, the activity for each neuron is recorded and denoted by a $1$ (spike) or a $0$ (no spike), resulting in a binary vector of size 160 for each time frame. Each iteration of the movie consists of $953$ timeframes.

We selected subsets of the ganglion cells with high mutual information (MI). This selection procedure is described in detail in Appendix~\ref{app:data}. To obtain a representative training set, the data is divided into iterations of the movie. The first $30$ iterations are selected for the training set $\mathcal{D}_{\text{train}}$, resulting in a set of $28590$ binary vectors. Other iterations of the movie are grouped per $10$ to form test sets, resulting in 26 separate test sets. Consequently, all timeframes occur in an equal ratio within the training set and the test sets. Since many test sets are available, this provides an excellent method for testing how well the QBM generalizes to unseen data.

\begin{figure}[ht]
    \centering
    \includegraphics[width=0.9\linewidth]{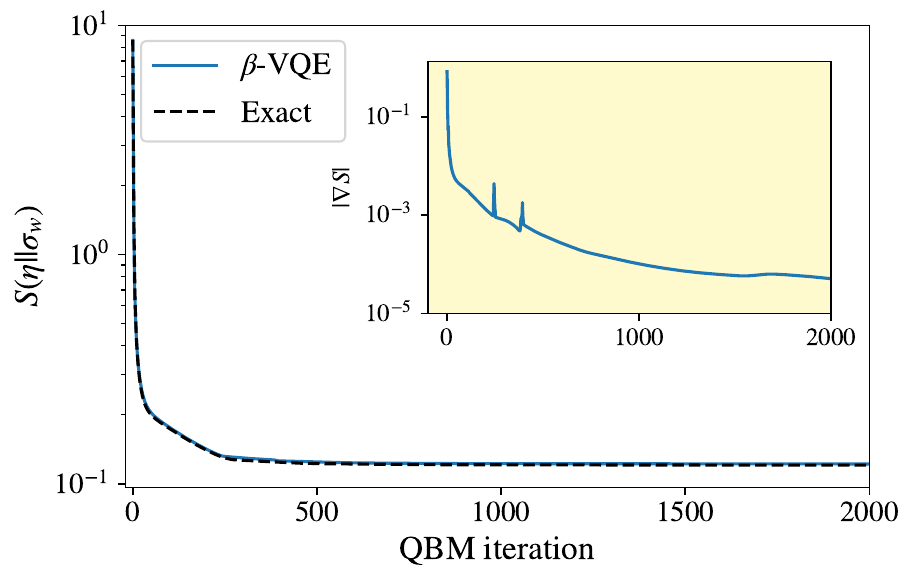}
    \caption{Convergence of the QBM training on the training set $\mathcal{D}_\text{train}$ of the classical salamander retina data of size $n=8$ using rank $R=16$ and depth $d=8$. The nested-loop algorithm closely approximates exact QBM gradient optimization in this setting. The inlay shows the size of the gradient. The QBM achieves a quantum relative entropy of $S=0.12$. A classical relative entropy, or KL-divergence, of $0.010$ is achieved on the training set, which is significantly lower than the classical BM that achieves a KL-divergence of $0.022$. The corresponding fidelity for the QBM is $F=0.998$, compared to $F=0.269$ for the classical BM.}
    \label{fig:convergence_salamander}
\end{figure}

In Fig.~\ref{fig:convergence_salamander}, we show the convergence of the quantum relative entropy during training. We observe that $S(\eta\|\sigma_w)$ decreases monotonically and saturates at $S=0.12$. This is close to the value obtained from learning with the exact gradient (dashed line). To compare these results to classical models, the QBM probability distribution over classical states can be obtained through projection, i.e. $p_{\text{QBM}}(\bm{s}^i) = \Tr(\outerproduct{\bm{s}^i}{\bm{s}^i} \sigma_w)$. We can then directly compare the classical relative entropy, or KL-divergence, to the classical BM. The QBM achieved a KL-divergence of $0.010$ on the training data. In contrast, the classical BM only achieves a KL-divergence of $0.022$ when fully converged. These results generalize well to unseen data. The QBM outperforms the classical BM, achieving a lower KL-divergence on all test sets. A histogram of the results is shown in Appendix~\ref{app:data}.

Note that the norm of the gradient, shown in the inset of Fig.~\ref{fig:convergence_salamander}, is $\mathcal{O}(10^{-4})$, which is still significantly nonzero. In contrast, the classical BM can be trained until the norm of the gradient reaches machine precision. The convergence of the QBM was numerically studied in~\cite{Kappen_2020} as a function of the rank of the target state $\eta$ where it was shown that convergences deteriorates with lower rank. It has been shown that later training iterations tend to increase the magnitude of the weights, akin to lowering the effective temperature of the QBM. During this phase, the ground state of the QBM, $\psi_{w}^{0}$, already accurately represents the target state despite the nonzero gradient, achieving better results than the classical BM. This is also reflected by the high fidelity of $F=0.998$ between the ground state and the target.

\begin{figure}[t]
    \centering
    \includegraphics[width=0.9\linewidth]{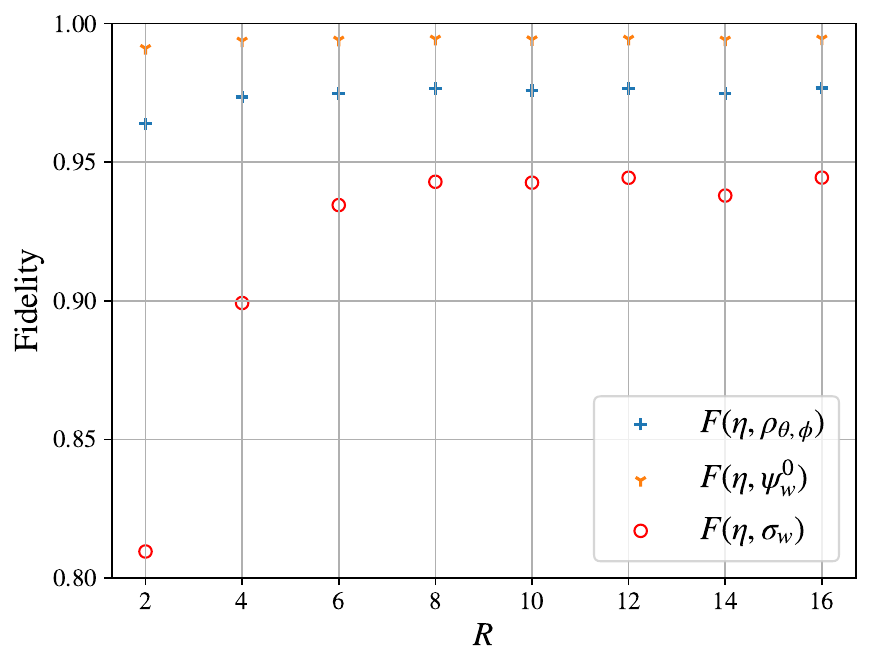}
    \caption{Fidelity achieved with different $\beta$-VQE rank $R$ and depth $d=n$ on salamander retina data of size $n=8$. The $\beta$-VQE achieves fidelities $>0.96$ even with small $R$. While the full QBM state $\sigma_w$ is still mixed resulting in a lower fidelity, the ground state $\psi_{w}^{0}$ achieves a fidelity of $>0.99$ in all cases. $\sigma_w$ can be turned into a pure state by multiplying the weights by a large constant factor.}\label{fig:SamplesSalamander}
\end{figure}

We investigate some important aspects of the learning algorithm, starting with the rank used in the truncated-rank $\beta$-VQE inner loop. In Fig.~\ref{fig:SamplesSalamander} we show the fidelity for different states obtained at the end of our algorithm. The fidelity between the QBM and $\eta$, indicated by the blue crosses, increases substantially with increased $\beta$-VQE rank, even though the target is a pure state. This might seem counter-intuitive but is a direct result of the low-rank approximation of the gradient. For a very low rank $\beta$-VQE, the approximation to the gradient becomes vanishingly small once the ground state accurately approximates the target pure state. During that stage, the exact $\sigma_{w}$ is still very mixed, but the gradient is approximated as if it were close to a pure state. Conversely, when a higher rank approximation is used, the gradient is still nonzero even if the ground state of $\sigma_w$ closely resembles the target pure state, allowing the QBM to converge further towards a pure state. The fidelity between the target and the ground state of the QBM is depicted in Fig.~\ref{fig:SamplesSalamander} by the yellow triangles. High fidelity of $F>0.99$ is achieved with as few as two samples.

\begin{figure}[t]
    \centering
    \includegraphics[width=.9\linewidth]{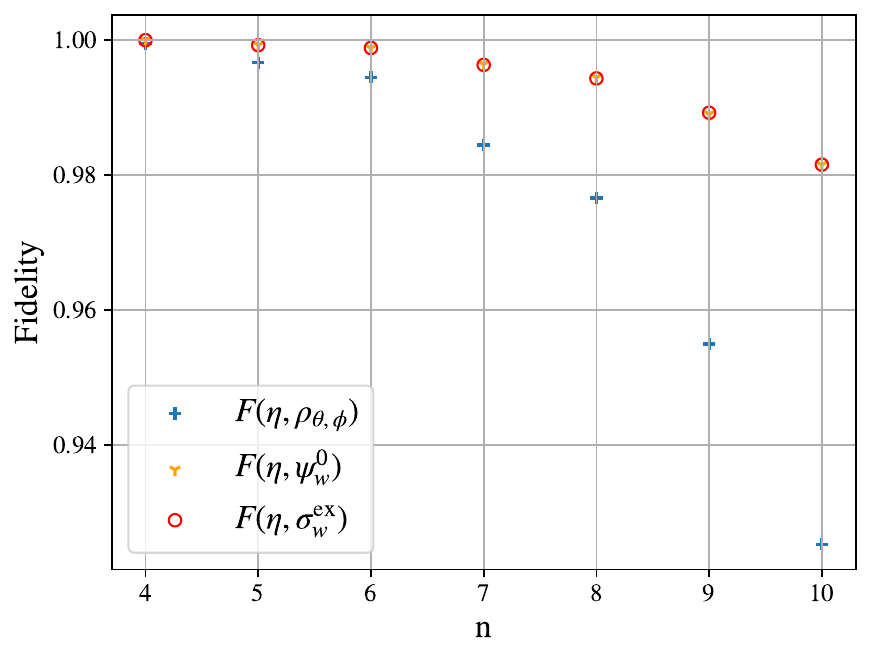}
    \caption{Fidelity for different system sizes using rank $R=2n$ and depth $d=n$. For all sizes, a fidelity of $F>0.98$ is achieved between the target state and the ground state of the QBM. The fidelity drops for larger system sizes, which is due to model mismatch between the QBM and the target. The error on the fidelity introduced by the $\beta$-VQE is of order $\mathcal{O}(10^{-4})$.}
    \label{fig:salamander_size}
\end{figure}

Another important aspect is the scaling of the performance of the algorithm with the number of features (qubits) $n$. In order to investigate this, we train the QBM on subsets of the salamander retina data of different sizes. We set the circuit depth equal to the number of features, which was heuristically found to be a good rule of thumb, see Appendix~\ref{app:depth}. Motivated by the previous sampling analysis, we also scale the rank of $\beta$-VQE linearly with $n$. In Fig.~\ref{fig:salamander_size} we show the fidelity obtained with these heuristic guidelines. The QBM trained by the nested-loop algorithm is able to obtain a fidelity of $F>0.98$ for all system sizes studied here. Note that the fidelity decreases for larger systems. This is because of a model mismatch between the QBM and the target, as this drop is observed also in the exact QBM. A more expressive Hamiltonian ansatz, e.g., with the inclusion of third-order interaction terms, might yield more accurate results for larger system sizes.

\subsection{Quantum data results}
\begin{figure}[t]
    \centering
    \includegraphics[width=\linewidth]{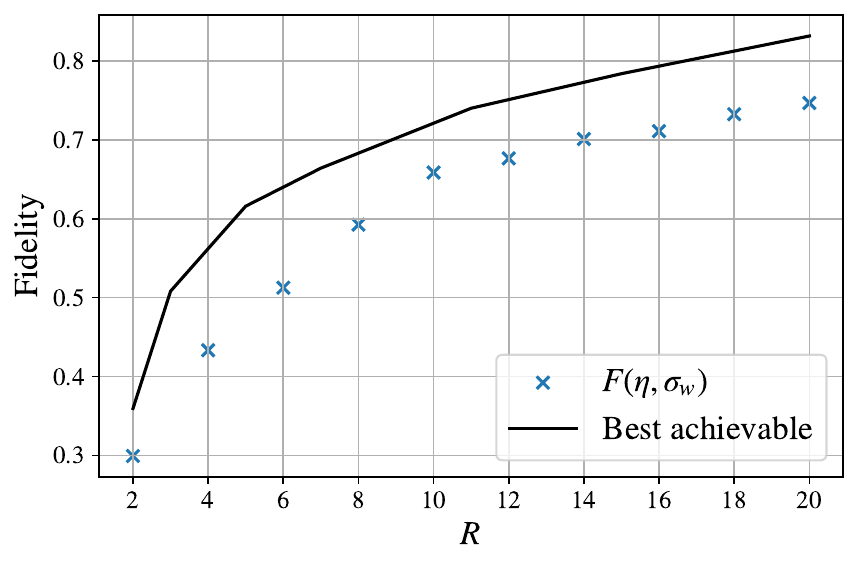}
    \caption{Fidelity achieved between the target XXZ model $\eta$ of size $n=8$ and the QBM using the nested-loop algorithm, as a function of the rank of the $\beta$-VQE using depth $d=n$. The black line represents the fidelity between the model and an exact fixed rank approximation, and as such represents the best achievable fidelity for the corresponding fixed rank $\beta$-VQE.}
    \label{fig:xxz_samples}
\end{figure}

In our second example, we train a QBM to model quantum data. As target state $\eta$ we take the finite temperature Gibbs state $\eta=e^{-\beta H_{\mathrm{XXZ}}}/Z$ with 
\begin{equation}
\label{eq:H_xxz}
H_{\mathrm{XXZ}} = \sum_{i=1}^{n-1}J\left(\sigma^{x}_i\sigma^{x}_{i+1}+\sigma^y_i\sigma^y_{i+1}\right)+\Delta \sigma_i^z\sigma^z_{i+1},
\end{equation} 
with $J=-1$, $\Delta=-\frac{1}{2}$, and $\beta=1$.
Quantum data taken from systems at finite temperatures is usually characterized by a high rank. However, we show that the $\beta$-VQE can be used to get an efficient low-rank approximation to the target state. We train the QBM with the nested-loop algorithm, again analyzing the effect of the rank. The results for the quantum data are shown in Fig.~\ref{fig:xxz_samples}. In contrast to the results for classical data shown in Fig.~\ref{fig:SamplesSalamander}, we require a higher rank approximation to obtain a high fidelity. By increasing the rank, the $\beta$-VQE is more capable of representing the mixed QBM state. In the truncated-rank $\beta$-VQE, one has the freedom to increase the rank freely during training, at the cost of a higher computational demand. For mixed target states, it can be efficient to start out with a low-rank approximation for early iterations of the outer loop, when only a rough approximation of the QBM gradient is sufficient for training, and increase the rank as the gradient approaches zero.

\subsection{Training QBM with the nested loop in practice}
\label{sec:practice}
All the simulations in previous sections assume access to the exact quantum state. This is possible via classical statevector simulations which are intractable in general. To overcome this, one can execute the required circuits on a quantum computer. This introduces two new sources of error. First, the quantum computer is affected by various sources of hardware noise~\cite{Preskill_2018,nielsen_2001}. While little information is available on the noise for a specific quantum chip, we believe this noise is biased in general. Thus, this is a source of error that cannot be completely mitigated by taking many measurements. Second, the output statistics are approximated from a finite number of shots (measurements). This is important, for example, when estimating the gradient of $\beta$-VQE in Eq.~\eqref{eq:GradTheta}. When hardware noise is absent, the error due to a finite number of shots ($M$) scales as $\epsilon \propto \frac{1}{\sqrt{M}}$. In this section, we take a closer look at these sources of errors in the context of QBM training.

We use Qiskit~\cite{Qiskit} to simulate both the noise-free and noisy hardware. For the noisy hardware simulations, we use the \emph{fake provider} module from Qiskit to mimic the behavior of IBM's quantum devices. We use a subset of the salamander retina data of size $n=4$ and the Hamiltonian ansatz given in Eq.~\eqref{eq:H_qbm}. For $\beta$-VQE we truncate the rank to $R=2$ and use the circuit in Fig.~\ref{fig:circuit} with a depth of $d=2$.

\begin{figure}
    \centering
    \includegraphics[width=7.85cm]{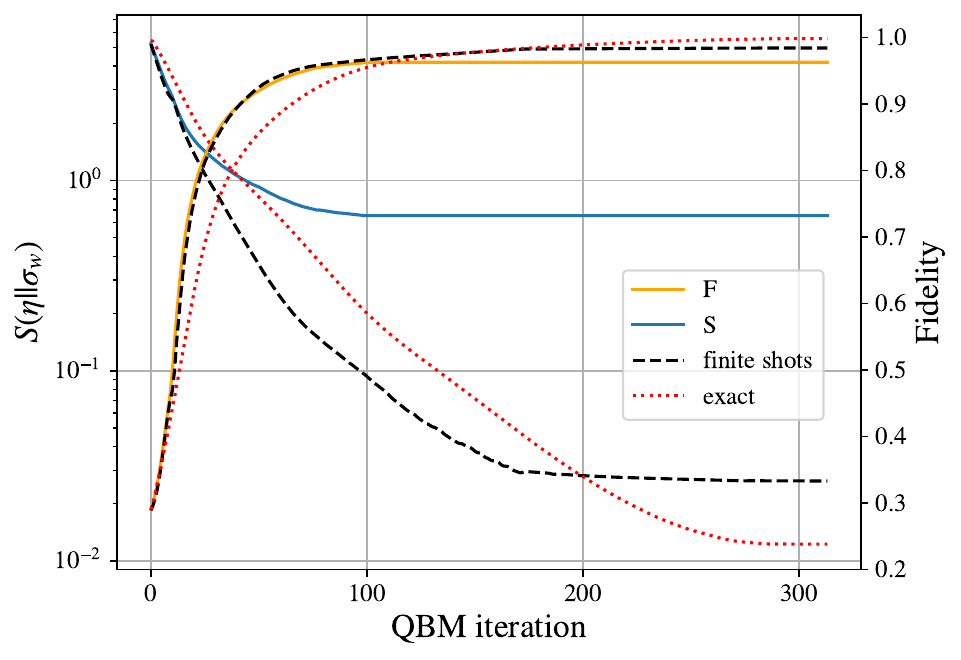} 
    \caption{Quantum relative entropy and fidelity during QBM training on the classical salamander retina data target of size $n=4$ by using exact expectation values (red doted line) and using a noise-free (dashed black line) and a noisy (solid lines) quantum simulator with finite shots to approximate the expectation values. We use the truncated  $\beta$-VQE with $R=2$ and $d=2$. To produce noisy hardware results, Qiskit's FakeKolkata backend is used, which mimics the noisy behavior of IBM's \emph{ibmq\_kolkata} device. We use $5000$ shots for approximating the expectation values present in estimating the variational free energy, and its gradient using the parameter shift rule. While there is a significant difference between relative entropies, the fidelities are very similar. A fidelity of $0.983$ and $0.962$ is achieved in noise-free and noisy hardware cases with finite shots respectively. When using exact expectation values, a fidelity of $0.998$ is achieved.
    }
    \label{fig:S_F_noise_free_noisy}
\end{figure}

Figure~\ref{fig:S_F_noise_free_noisy} compares the QBM training in three different settings in terms of estimating the expectation values: using exact values, using a noise-free quantum simulator with $5000$ shots, and finally using a noisy quantum hardware simulator with $5000$ shots. The FakeKolkata backend from Qiskit, which mimics the behavior of IBM's \emph{ibmq\_kolkata} device, is used to produce the noisy hardware results.

\begin{figure}[t]
    \centering
    \includegraphics[width=7.85cm]{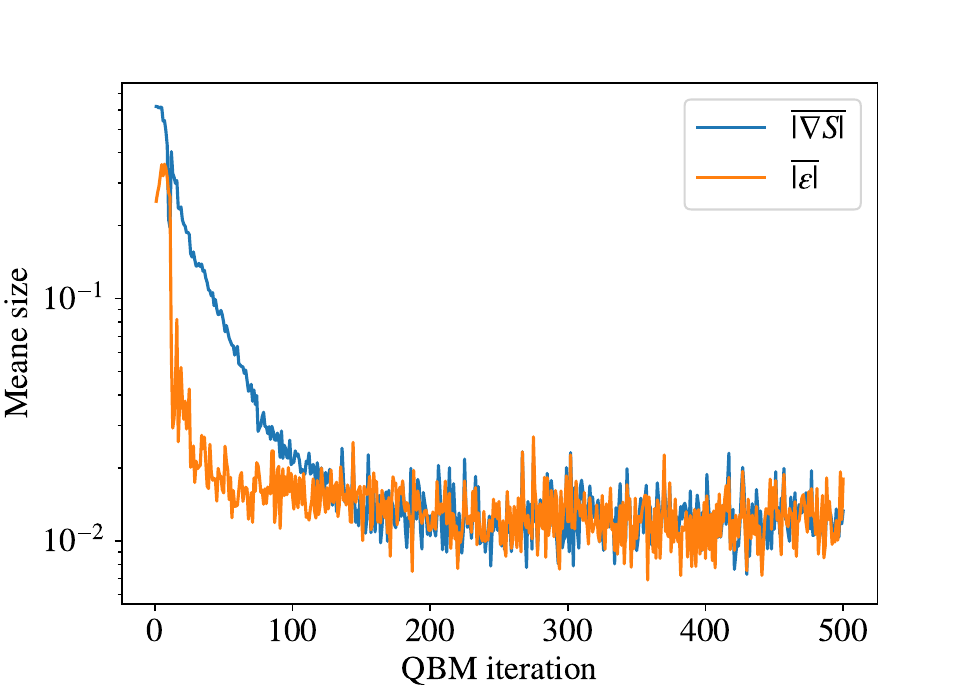} 
    \\
    \includegraphics[width=7.9cm]{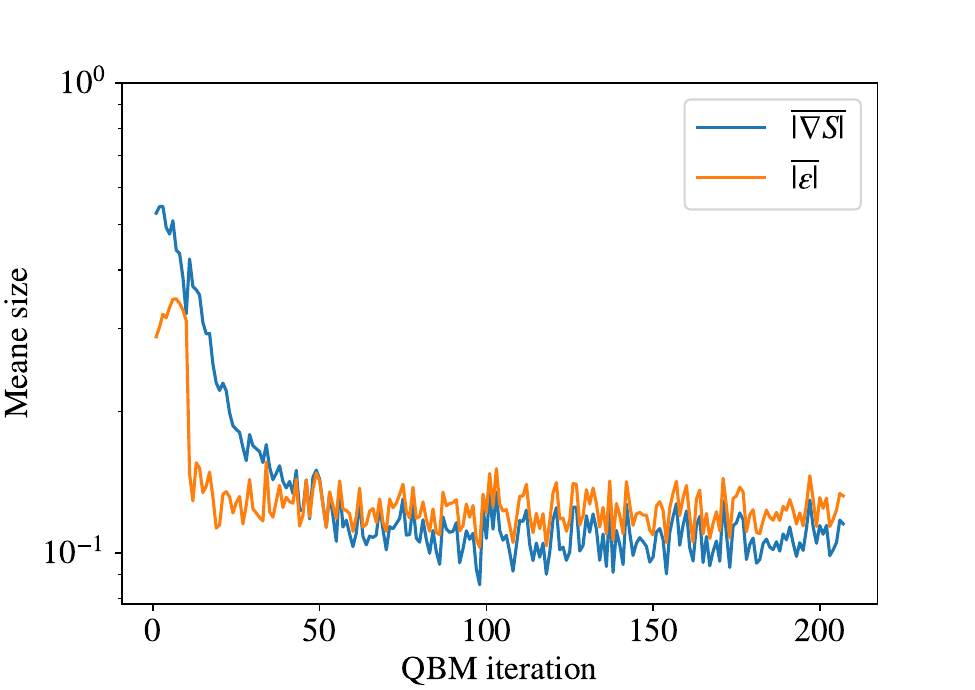}
    \caption{The mean size $\overline{|\nabla S|}$ of the QBM gradient (blue) and the mean size of the error $\epsilon$ introduced by the $\beta$-VQE approximation (orange) during the training in the noise-free (top panel) and noisy (bottom panel) quantum simulators. The training process and model ansatz here is exactly the same process that we have used to make the plots for noise-free and noisy quantum simulators with finite shots in Fig.~\ref{fig:S_F_noise_free_noisy}.
    The QBM converges until the error introduced by the $\beta$-VQE becomes the dominant factor of the gradient. The error decreases initially but finally fluctuates around $\frac{1}{\sqrt{5000}} \approx 0.014$ in the noise-free quantum simulator case as expected. In the noisy quantum simulator case the error $\epsilon$ is larger, which forces the training to stop earlier than the noise-free quantum simulator case. The resulting QBM has a higher quantum relative entropy as shown in~Fig.~\ref{fig:S_F_noise_free_noisy}.
    }
    \label{fig:mean_DS_bvqe_error}
\end{figure}

As expected, the learning stops earlier in the presence of noise. An increase in the noise causes the errors introduced by the $\beta$-VQE algorithm to reach the size of the QBM gradient faster. 
The size of the error compared to the size of the gradient is shown in~Fig.~\ref{fig:mean_DS_bvqe_error} for the noisy and noise-free quantum simulators, when both are using the same number of shots to approximate the expectation values. 
As expected, a higher (lower) quantum relative entropy (fidelity) is achieved using a noisy hardware compared to a noise-free one.

\begin{figure}[t]
    \centering
    \includegraphics[width=7.85cm]{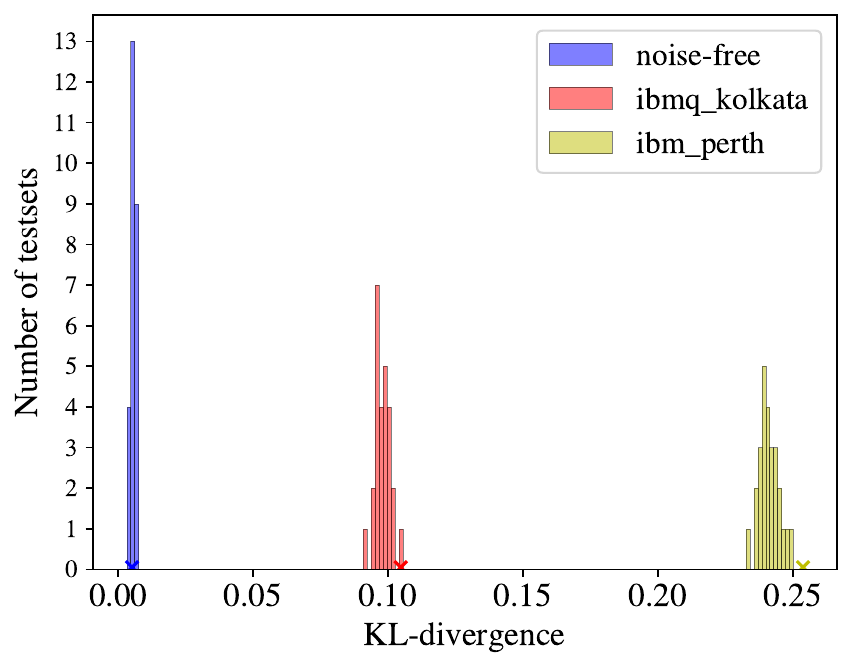} 
    \caption{KL divergence to the target salamander retina data for probability distributions obtained from real noisy quantum hardware and noise-free quantum simulator. We use a system size of $n=4$ and a rank $R=2$ and depth $d=2$ $\beta$-VQE. In the real quantum hardware cases, the resulting $\beta$-VQE is implemented on the physical quantum devices \emph{ibmq\_kolkata} and \emph{ibm\_perth}, and the final probability distributions is obtained using sampling. 
    As it is clear from the figure, sampling from the physical devices results in a higher KL divergence compared to the noise-free quantum simulator case due to the noise present in the current quantum devices.
    Parameters for \emph{ibmq\_kolkata} and \emph{ibm\_perth} are obtained by simulating the QBM training with the FakeKolkata and FakePerth backend respectively.
    The crosses denote the KL divergence on the training set, and the histogram shows the KL divergence on test sets.}    
\label{fig:kls_kolkata.pdf}
\end{figure}

Finally, we deploy the QBM model trained using the FakeKolkata backend on the corresponding real quantum hardware device, \emph{ibmq\_kolkata}. We sample the thermal state in the computational basis and compare its KL-divergence to the target distribution. The target distribution is resampled a number of times in order to obtain different test sets. The results are shown in Fig.~\ref{fig:kls_kolkata.pdf}. As expected, the \emph{ibmq\_kolkata} (red bars) yields KL-divergences higher than those of the noise-free quantum simulator (blue bar). However, deploying the same QBM model on another IBM device, \emph{ibmq\_perth} (yellow bars), results in a KL-divergence about twice as high as using \emph{ibmq\_kolkata}, demonstrating the recent improvements made to noise reduction in quantum hardware. We expect that further advancements in noise reduction will contribute significantly to improving these results. Note that QBM models trained in the noisy setting are to some extent bound to the specific hardware~\cite{Benedetti_2017} (data not shown).
Further investigation is needed to examine how hardware noise affects QBM training by increasing the number of qubits.

\section{Discussion and Conclusion}
\label{sec:concl}
Training QBMs by minimizing the quantum relative entropy with gradient descent requires calculating the difference in statistics between the target and the model. In general, this is intractable because it involves computing exact Gibbs state expectation values. This has been shown to be QMA-hard~\cite{Bravyi_2021}, which means that even a quantum algorithm could take exponential time. We show that the $\beta$-VQE can be used as an efficient heuristic for obtaining the gradient in practice, closely approximating QBM training with the exact gradient. When combined with the QBM training loop, we dubbed this the nested-loop training algorithm. 

The QBM trained with the nested-loop algorithm outperforms the classical BM, achieving a significantly lower KL-divergence. In comparing the two models, it should be noted that the QBM has about three times as much parameters as the classical BM. However, the classical BM is a subset of the QBM using only diagonal $\sigma^z$ terms, and as such can only represent a diagonal state regardless of the amount of parameters used. Furthermore, note that the results in this paper are indicative of the improvements that the QBM can achieve over classical models, and do not necessarily represent the best possible results. Both the QBM and the BM results can be improved by adding hidden variables~\cite{Wiebe_2019,Zhang_2018,Carleo_2017}, at a higher computational cost.

When the target is constructed from classical data, we find that the QBM can be trained using a low-rank version of $\beta$-VQE. This is also observed for quantum data originating from low-temperature systems. For high-temperature quantum systems, a high rank is needed to achieve high fidelity. Yet, for the data considered here, we find numerically that a circuit depth scaling linearly with the system size is sufficient. We also find that a warm-start strategy for $\beta$-VQE can significantly reduce the overall computational cost. The nested-loop algorithm provides a method for training the QBM on near-term quantum devices. 

The nested-loop algorithm is a synergy between two models, where the advantages of one model compensate for the weakness of the other. Having two model density matrices, $\sigma_w$ and $\rho_{\theta, \phi}$, may seem redundant, and one may wonder why the $\beta$-VQE is not used to represent the data directly. However, optimization of the $\beta$-VQE on the data density matrix directly requires computing the fidelity, and hence the overlaps between quantum states. This is intractable for large system sizes. The QBM serves as an intermediary, providing a local parameterized Hamiltonian that can be evaluated on a circuit. Similarly, the $\beta$-VQE provides the low-rank approximation necessary to train the QBM. Previous experiments with rank-one approximations for QBM training have revealed that this leads to problematic level crossings that prohibit convergence, even in the case where the target distribution can be well approximated with a rank-one density matrix. An example is shown in Appendix~\ref{app:rank1}. Thus, even for a rank-one target, it is often necessary to use higher-rank approximations for the QBM to converge. The $\beta$-VQE provides a useful approximation where the rank can be freely chosen, and whose statistics can be efficiently computed by evaluation of a quantum circuit.

At the end of training, the user is left with two density matrix representations of the target state. Depending on the purpose, either one can be used as a final approximation to the data. For quantum tomography the QBM state $\sigma_w$ is most useful, revealing a possible candidate for the underlying Hamiltonian. For classification and generative tasks, one might directly use the $\beta$-VQE density $\rho_{\theta, \phi}$.

Alternative methods for approximating Gibbs states could act as an inner loop instead of the $\beta$-VQE~\cite{Liu_2021}. Recently, a new method for generating Gibbs states was introduced called the Noise-Assisted Variational Quantum Thermalization~\cite{Foldager_2022} (NAVQT), where noisy quantum gates thermalize a pure state. While their idea is promising, the $\beta$-VQE achieved better results on models with randomized parameters. Several other approaches rely on preparing a purification on a larger system~\cite{Wang_2021, Consiglio_2023}, including the thermofield double (TFD) state~\cite{Wu_2019, Martyn_2019}. The Gibbs state is obtained from the pure state by tracing out the additional subsystem. Evidently, this demands a significant increase in the number of qubits needed, up to doubling the system size for the TFD. This computational demand makes it less suitable for the QBM inner loop that has to be evaluated many times during training. In contrast, the $\beta$-VQE is initiated with a mixture of classical states. For target systems that are not very mixed, the truncated $\beta$-VQE needs only a few states to obtain an accurate approximation, especially for a pure target state. Nevertheless, purification methods could be preferable for high-temperature target systems that are close to maximally mixed. For such systems, the $\beta$-VQE needs to sample over a significant portion of the Hilbert space to obtain accurate results, whereas purification methods still need only a single state in a system size twice as large.
Finally, the classical networks considered in $\beta$-VQE provide a more tractable and expressive latent space than the product state ansatzes considered in earlier work~\cite{Verdon_2019,Martyn_2019}, making it suitable for QBM training.

\section*{Data availability statement}
No new data were created or analysed in this study.

\section*{Acknowledgments}
We thank Samuel Duffield and Kirill Plekhanov for fruitful discussions. We thank Manu Compen for earlier collaborations on rank-one QBM training. This publication is part of the `Quantum Inspire – the Dutch Quantum Computer in the Cloud' project (NWA.1292.19.194) of the NWA research program `Research on Routes by Consortia (ORC)', which is funded by the Netherlands Organization for Scientific Research (NWO).

\bibliography{refs}

\appendix

\section{Salamander retina data}
\label{app:data}
For training on the classical salamander retina data, we select subsets of various sizes $n$ with high MI. We initialize the dataset by selecting the pair of neurons that have the highest pairwise MI between all neuron pairs. Subsequently, the neuron that has the highest MI with the existing subset is selected and added. This is repeated until a dataset of the desired size is constructed.

\begin{figure}[h]
    \centering 
    \includegraphics[width=8.25cm]{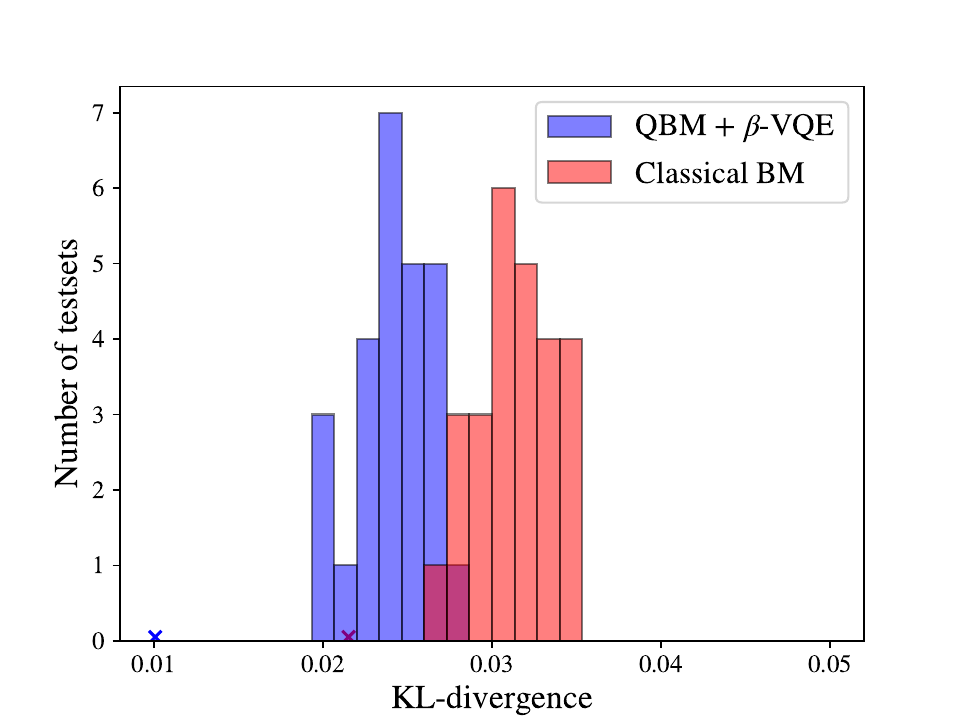}
    \caption{Histogram of KL-divergences for both the QBM and classical BM on the salamander retina data of size $n=8$. We used a $\beta$-VQE of rank $R=16$ and depth $d=8$. The crosses denote the results on the training set and the histograms show the results on the test sets. The nested-loop QBM outperforms the classical BM, achieving a lower KL-divergence on all test sets.}
    \label{fig:retdat}
\end{figure}

As shown in~\cite{Kappen_2020}, the QBM with exact gradient computation outperforms the classical BM on this data set. We reproduce these results for the QBM trained with the nested-loop algorithm. The QBM achieves a lower KL-divergence on all test sets. A histogram of these results is shown in Figure~\ref{fig:retdat}.

\section{Circuit-depth analysis}
\label{app:depth}

In this section, we assume the QBM in the outer loop is fixed to the Heisenberg XXZ model in Eq.~\eqref{eq:H_xxz} with $\beta = 1$. We then analyze the effect of circuit depth in $\beta$-VQE which, as a reminder, is executed in the inner loop to approximate the QBM.

\begin{figure}[h]
    \centering    
    \includegraphics[width=8cm]{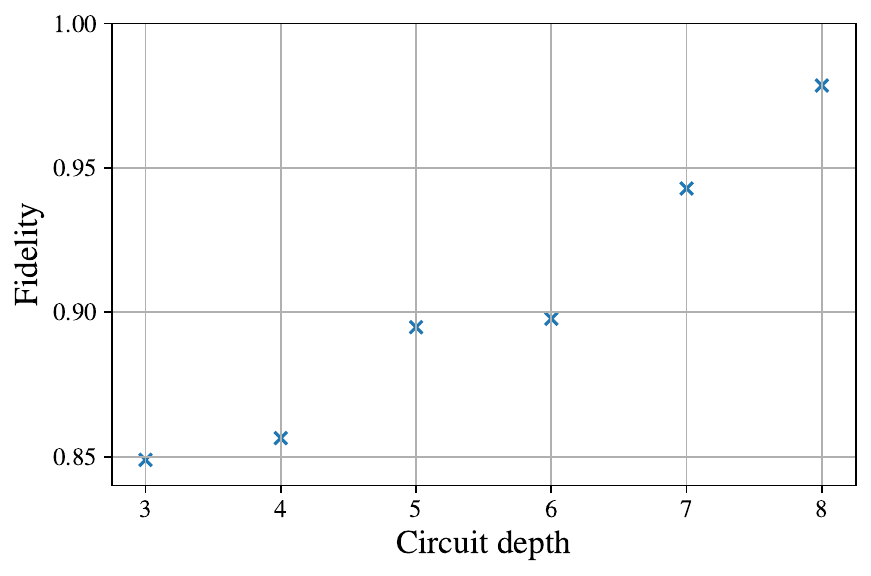}
    \vspace{-.25cm}
    \caption{Fidelity between the QBM and the target Heisenberg XXZ model in Eq~\ref{eq:H_xxz} of size $n=8$ for different circuit depths. A rank $R=16$ $\beta$-VQE was used.}
    \label{fig:f_vs_depth}
\end{figure}

We use the exact gradient in Eq.~\eqref{eq:GradTheta} and increase the circuit depth from $3$ to $8$ layers. We compute the Uhlmann-Jozsa fidelity in Eq.~\eqref{eq:fidelity} between the $\beta$-VQE and the QBM. The results are shown in Figure~\ref{fig:f_vs_depth}. Clearly, deeper circuits can achieve higher fidelity. We emphasize however that increasing the circuit depth is computationally expensive as the resulting circuits contain more parameters to train. A circuit depth of $8$ is sufficient to obtain $F>0.98$. 

\vspace{-.5cm}

\section{Rank-one QBM training}
\label{app:rank1}

\begin{figure}[h]
    \centering
    \includegraphics[width=8.4cm]{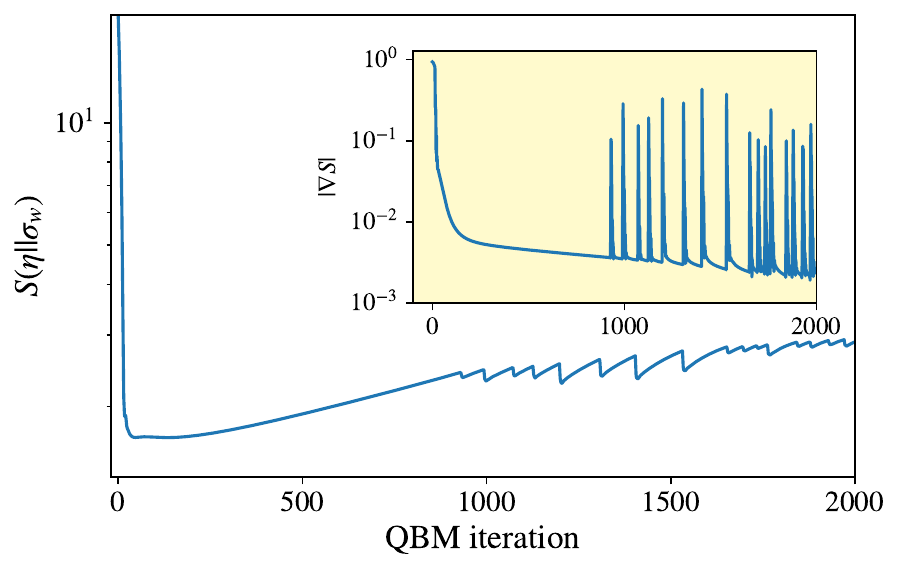}
    \vspace{-.25cm}
    \caption{Convergence of the QBM trained using a rank-one approximation on salamander retina data of size $n=8$. We used a $\beta$-VQE of rank $R=1$ and depth $d=8$. Problematic level crossings prevent further convergence.}
    \label{fig:rank1_convergence}
\end{figure}

\begin{figure}[h]
    \centering
    \includegraphics[width=7.5cm]{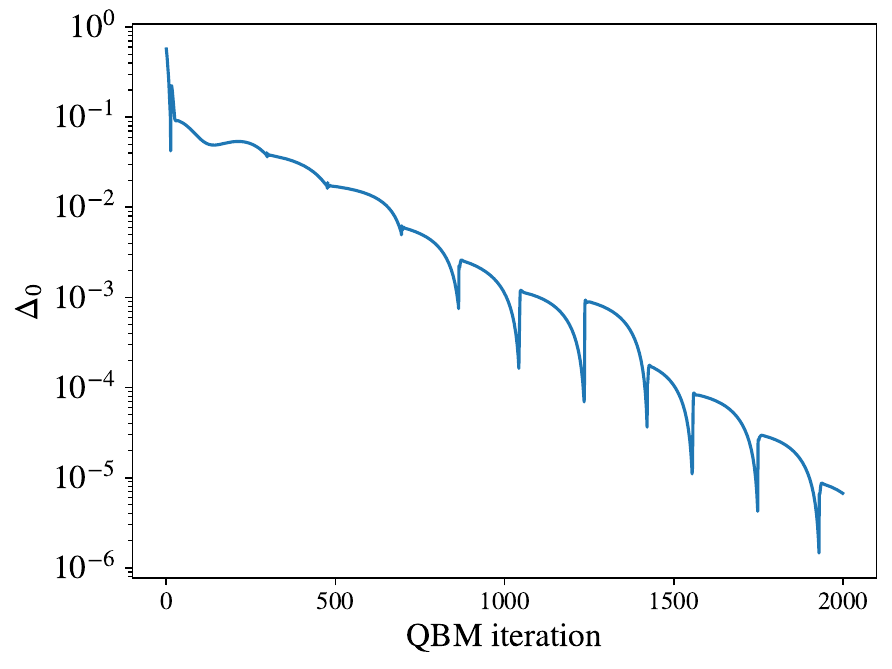} \\
    \includegraphics[width=7.4cm]{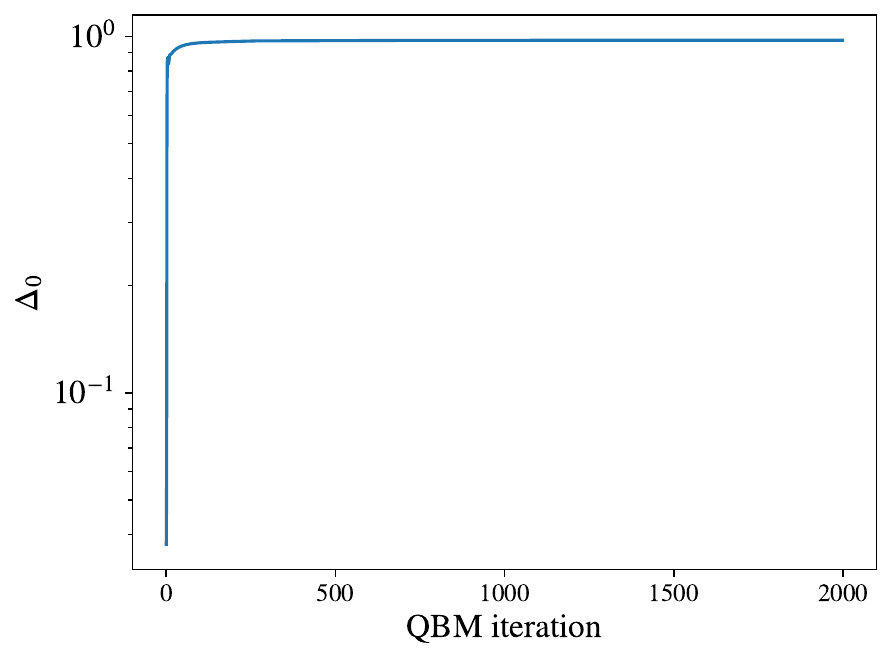}
    \caption{Top: Spectral gap of the QBM during training on salamander retina target state of size $n=8$ using a rank $R=1$ approximation of depth $d=8$. After a level crossing is passed, the QBM converges towards the point of degeneracy
    Bottom: Spectral gap of the QBM during training on the same salamander retina target state using a rank $R=16$ depth $d=8$ $\beta$-VQE in the nested-loop algorithm. The QBM now smoothly converges.
    }
    \label{fig:spectral_gap}
\end{figure}

For targets encoding classical data, the QBM usually converges to a rank-one solution. Thus, one may wonder if it is possible to train the QBM using only the ground state to approximate the gradient. This would provide an opening to using various methods of ground state approximation. 
We find that using a rank-one approximation does not work in general due to problematic level crossings. As the QBM converges, level crossings between the ground state and the first excited state can occur. Consequentially, the ground state approximation of the gradient changes drastically between two iterations. This is troublesome if the gradient after the level crossing again directs the model back over the level crossing. In such a case, one can only converge closer to the exact point of degeneracy, and at that point, it is not possible to train the QBM further using the ground state approximation. Such a case is plotted in Figs.~\ref{fig:rank1_convergence} and~\ref{fig:spectral_gap}. 

One might see the resemblance between the spikes in the norm of the gradient in Fig.~\ref{fig:rank1_convergence}~(inset) and the two spikes in Fig.~\ref{fig:convergence_salamander}~(inset). However, they are due to different events. For rank-one training in Fig.~\ref{fig:rank1_convergence} the spikes coincide with the level crossings shown in in Fig.~\ref{fig:spectral_gap}. Instead, for the truncated-rank $\beta$-VQE the system is always gapped. In this case, the spikes are due to the high momentum and variable learning rate that we use during training. These settings accelerate training, especially for rank-one targets. But, sometimes, they cause the QBM to overshoot the optimal point, causing the spikes.

\FloatBarrier

\end{document}